# *Geometry-Invariant Resonant Cavities*

**Authors:** I. Liberal[1,2], A. M. Mahmoud[2] and N. Engheta[2*]

**Affiliations:**

[1]Electrical and Electronic Engineering Department, Universidad Pública de Navarra, Pamplona E31006, Spain.

[2]Department of Electrical and Systems Engineering, University of Pennsylvania, Philadelphia, Pennsylvania 19104, USA.

*Correspondence to: engheta@ee.upenn.edu

## Abstract

Resonant cavities are one of the basic building blocks in various disciplines of science and technology, with numerous applications ranging from abstract theoretical modeling to everyday life devices. The eigenfrequencies of conventional cavities are a function of its geometry, and, thus, the size and shape of a resonant cavity is selected in order to operate at a specific frequency. Here, we demonstrate theoretically the existence of geometry-invariant resonant cavities, i.e., resonators whose eigenfrequency is invariant with respect to geometrical deformations. This effect is obtained by exploiting the unusual properties of zero-index metamaterials, which enable decoupling of the time and spatial field variations. This new class of resonators may inspire alternative design concepts, and it might lead to the first generation of deformable resonant devices.

**Main Text:**

The dynamics of many physical systems are usually described in terms of wave equations subjected to certain boundary conditions. This is the case, for example, in classical and quantum mechanics, electromagnetics, acoustics, fluid dynamics, etc. Specifically, when considering source-free time-harmonic $\exp(-i\omega t)$ fields, one finds that the solutions to these equations, subject to specific boundary conditions, often take place at specific discrete $\omega$-frequency values, usually labeled as eigenfrequencies, or resonant frequencies (*1*). In general, wave equations interrelate both spatial and time variations of the fields (e.g., consider the vector wave equation of the electric field in classical electromagnetics: $\nabla \times \nabla \times \boldsymbol{E} + c^{-2}\partial_t^2 \boldsymbol{E} = \boldsymbol{0}$ (*2*)). Consequently, eigenfrequencies are determined by the geometry at hand.

This fundamental principle shapes the way we address various phenomena and develop technology. In fact, researchers, engineers and designers across a myriad of fields stretch their minds to find the appropriate geometry to operate at a specific frequency. Furthermore, fabrication imperfections degrade the performance of the devices, as well as hinder the application of thrilling physical concepts that may unfortunately require too stringent fabrication tolerances. Therefore, we could wonder if, as it is symbolically sketched in Fig. 1, it could be possible to find scenarios in which the eigenfrequencies of a resonant cavity are invariant with respect to geometrical transformations. If so, this would represent a complete change in the mindset behind design processes, and it might open up the possibility of developing resonant devices that continue working even under severe geometrical deformations with interesting applications, for example, in tailoring light-matter interaction and quantum emission in such deformable structures.

Naturally, the idea of a geometry-invariant resonator challenges our intuition on how waves usually behave. However, the fields of topological insulators (*3*, *4*) and topological photonics (*5–10*) have revealed that certain physical quantities are preserved under continuous deformations. Moreover, during the past several years metamaterials have demonstrated that waves can be manipulated in unconventional manners (*11–20*). For instance, metamaterials featuring extreme parameters, such as epsilon-and-mu-near-zero (EMNZ) and zero refractive index structures, have been found to support fields with static spatial distributions, while maintaining their temporarily dynamic properties (*21–26*). This apparent decoupling between spatial and temporal domains encouraged us to believe that, indeed, resonators whose eigenfrequencies are invariant under geometrical transformations could be possible. Here, we analytically and numerically demonstrate that this is in fact the case, and that there are at least three distinct physical mechanisms in which zero index metamaterials enable the development of geometry-invariant resonant cavities.

In the following, we will concentrate on the classical source-free time-harmonic wave equation for the electric field $\boldsymbol{E}$ in nonmagnetic media: $\nabla \times \nabla \times \boldsymbol{E} - \varepsilon\,(\omega/c)^2 \boldsymbol{E} = \boldsymbol{0}$ (*2*), with $c$ being the speed of light in vacuum, and $\varepsilon$ the permittivity of the medium at hand. However, this must be considered only as a specific example of a more general concept that, as many other metamaterial paradigms, can be extrapolated to other forms of waves, such as acoustic, elastic, mechanical, and matter waves. Moreover, electromagnetic systems also represent an excellent test bench for future experimental verifications of the concepts introduced in this work. In fact, different experimental realizations of zero index electromagnetic metamaterials have already been reported in the form of naturally available materials (*27*, *28*), dispersion engineering in waveguides (*29*, *30*), photonic crystals (*25*) and artificial electromagnetic materials (*31*, *32*).

To begin with, one of the most striking properties of zero index metamaterials is their ability to "freeze" in space the phase and exceptionally also the magnitude variations of the fields (*21–24*). For instance, in epsilon-near-zero (ENZ) media – i.e., media whose relative permittivity is approximately zero, $\varepsilon \approx 0$ - the magnetic field parallel with the axis of a two-dimensional (2D) system must be uniform in order to avoid a singularity of the electric field $\boldsymbol{E} = i/(\omega\varepsilon)\nabla H_z \times \hat{\boldsymbol{z}}$ (*22–24*). One could anticipate that the influence of geometry is lessened in the presence of spatially uniform fields, since effectively the apparent wavelength in such media is very large. This intuition is indeed correct, and we show that uniform field distributions can be associated with 2D cavities whose eigenfrequency is invariant with respect to equi-areal transformations. To this end, let us consider, for example, a 2D cavity composed of an infinitely long dielectric circular cylinder of relative permittivity $\varepsilon_i$ and radius $r_i$ immersed in an ENZ host of arbitrary cross-sectional shape but area $A_h$ (see Fig. 2A), bounded by perfectly electric conducting (PEC) wall. As demonstrated in the supplementary online text, section 2.1, the eigenfrequencies obtained as solutions to the source-free electric field time-harmonic wave equation subject to the boundary condition $\hat{\boldsymbol{n}} \times \boldsymbol{E} = \boldsymbol{0}$ on the PEC wall are determined by the solutions to the following characteristic equation: $\omega = c \, 2\pi r_i / \left(A_h \sqrt{\varepsilon_i}\right) J_0'\left(\sqrt{\varepsilon_i} \, \frac{\omega}{c} \, r_i\right) / J_0\left(\sqrt{\varepsilon_i} \, \frac{\omega}{c} \, r_i\right)$, where $J_0(x)$ is the cylindrical Bessel function of the first kind and order zero, and $J_0'(x) = \partial_x J_0(x)$. This equation reveals that the eigenfrequencies of the eigenmode with uniform magnetic field in the ENZ host medium only depend on the characteristics of the internal particle $(\varepsilon_i, r_i)$, and the area of the ENZ host $A_h$. Therefore, the eigenfrequency is independent of the shape of this 2D cavity, as long as its cross-sectional area remains the same. As if it were an incompressible fluid, the cavity can be exposed to any equi-areal geometrical deformation while keeping the same eigenfrequency. Note that the set of allowed geometrical deformations also includes piercing - making holes - on the cavity. Therefore, the eigenfrenquency is immune even to evident changes in the topology of the cavity. A few examples of 2D cavities whose eigenfrequencies are the same and invariant with respect to equi-areal transformations are reported in Fig. 2B. These specific examples have been selected in order to illustrate the high degree of arbitrariness in the geometry of the cavities, including non-canonical shapes (Fig. 2B-I), different topologies (Fig. 2B-II), sharp corners (Fig. 2B-III), and high aspect ratios (Fig. 2B-IV). A more detailed description of the geometry of these cavities is gathered in Figs. S1-S4. Anticipating future experimental verifications of the presented results, the ENZ host has been modeled using silicon carbide (SiC) (*27*), whereas the internal dielectric cylinder is assumed to be silicon (Si), with relative permittivity $\varepsilon_i = 11.7$ (*33*). In this manner, our analysis includes the effect of the relatively high losses of SiC with relative value of the imaginary part of its permittivity to be 0.1 ($\varepsilon'' \approx 0.1$) in the vicinity of the SiC plasma frequency, $\omega_p = 2\pi \times 29.08 \times 10^{12}$ rad/s, where the real part of the relative permittivity is near zero ($\varepsilon' \approx 0$) (*27*). The radius of the cylinder ($r_i$=1.165 μm) and the area of the host ($A_h = 49\pi$ μm$^2$) have been selected such that the characteristic equation is satisfied at the SiC plasma frequency. The eigenfrequencies of these 2D resonators were computed numerically and are depicted in Fig. 2C. It is apparent from the figure that, despite their very distinct geometry, and despite the fact that realistic losses have been taken into account, the eigenfrequencies of these resonators deviate less than a 1.5% from the plasma frequency of SiC. Moreover, these small disagreements are only caused by the losses of SiC, which slightly deviate the response of the host from that of a pure ENZ medium. As a matter of fact, the eigenfrequencies converge even more closely to the SiC plasma frequency as losses decrease (see Fig. S5).

It is worth remarking that not all properties of the resonator are invariant with respect to geometrical deformations. For example, the quality factor $Q$ – i.e., the ratio between the energies stored and dissipated per cycle– strongly depends on the field intensity distributions in the resonator (*2*). Subsequently, as it is illustrated in Fig. 2C, deforming the cavities results in changes in the quality factor in excess of a 10%, while their resonance frequency stays effectively unchanged. In this manner, our calculations suggest the possibility of designing resonators with tunable quality factors, and, hence, adjustable decay rates, while keeping a constant resonant frequency. Moreover, Figs. 2C and S5 also serve to illustrate a unique property of the proposed geometry-invariant resonators that, to the best of our knowledge, has no counterpart in conventional resonators. Specifically, in a conventional resonator, the quality factor increases as losses decrease, and the eigenfrequency becomes more sensible against geometrical deformations. On the contrary, in our proposed geometry-invariant resonators, the smaller the losses the larger the quality factor, but also the more robust the eigenfrequency is against geometrical deformations (see Fig. S5). In this manner, the proposed idea enables the development of high $Q$ resonators whose eigenfrequencies are immune to geometrical deformations.

Even a more general invariance with respect to geometrical deformations may be found by noting that, as demonstrated in the supplementary online text, section 2.2, ENZ media may also support another modes as $\exp(-i\omega t)$ time-varying spatially "electrostatic" fields, which are different from what we discussed above. In other words, as the medium relative permittivity goes to zero, the Maxwell curl equation $\nabla \times \boldsymbol{H} = -i\omega\varepsilon_0\varepsilon\boldsymbol{E}$ may also support solutions with zero magnetic field $\boldsymbol{H} = \boldsymbol{0}$, but a non-zero and time-varying electric field, $i\omega\boldsymbol{E} \neq \boldsymbol{0}$. Naturally, the other Maxwell curl equation imposes that the associated electric field is irrotational $\nabla \times \boldsymbol{E} = i\omega\mu_0\boldsymbol{H} = \boldsymbol{0}$, since for this mode $\boldsymbol{H}$ is zero in the ENZ region. Thus, interestingly, ENZ media may support solutions to the wave equation in the form of spatially "electrostatic" distributions that are dynamically varying in time. Specifically, these spatially electrostatic field distributions are the solutions to Laplace equation in the ENZ host, subject to the appropriate boundary conditions. We emphasize that when the solution to the Laplace equation exists, the eigenfrequency plays no role in it as long as the ENZ host is still ENZ at that frequency. Therefore, if the boundary conditions on the ENZ host enable the existence of spatially electrostatic modes, then such a cavity has an eigenfrequency at the ENZ frequency, no matter what its geometry is. As shown in the supplementary online text, sections 2.3 and 2.4, the invariance of the eigenfrequency in the presence of time-harmonic spatially electrostatic fields can also be proven by using perturbational techniques. These modes can be excited in both 2D and 3D systems.

In order to illustrate this phenomenon with a specific example, let us consider a three-dimensional scenario in which a resonator is composed by a dielectric particle immersed in an ENZ host. In this case, since we want the magnetic field to be zero in the ENZ host medium, the boundary conditions impose that the tangential magnetic field on the surface of the dielectric particle must also vanish. Specifically, if the internal particle is a sphere, then this boundary condition is met at the solutions of the following characteristic equation: $\hat{J}_n \left( \sqrt{\varepsilon_i} \, \frac{\omega}{c} \, r_i \right) = 0$ for $n = 1,2, \ldots$ (see also supplementary online text, section 2.3).That is to say, the eigenfrequencies of the resonator correspond to the zeros of the functions $\hat{J}_n(x)$, representing the Schelkunoff form of the spherical Bessel functions of the first kind and order $n$ (2). Note that in this case there is not only one, but an infinite number of possible eigenmodes *n=1,2,3…* with geometry invariant properties. Moreover, due to the spherical symmetry of the internal particle, there are *2n+1* degenerate modes for each *n-th* eigenmode. We emphasize that the solutions to this characteristic equation only depend on the properties on the internal particle ($\varepsilon_i, r_i$), and are independent of the geometry of the main cavity. Therefore, the cavity can be of any size and shape, and it can also contain any number of holes in it. What is more, it can even be polluted with other particles with different dielectric materials, sharing the same ENZ host medium. In all these cases, the cavity will support an eigenmode at the ENZ frequency. This fact is illustrated in Fig. 3A, which shows four cavities with very distinct geometries, but that nevertheless share eigenmodes at the same eigenfrequency. Again, these specific cavities have been chosen to illustrate the high degree of arbitrariness in the geometry of the cavities (shape, topology, and in this case also size). A more detailed description of their geometry can be found in Figs. S6-S9. All cavities are composed of a SiC host containing a Si particle. In this case, a spherical particle of radius $r_i = 2.155$ $\mu$m has been selected to satisfy the characteristic equation, with $n = 1$, at the SiC plasma frequency. For the sake of brevity, Fig. 3A only depicts the electric field magnitude distribution of one of the three degenerate modes that can be excited in the vicinity of the SiC plasma frequency (each eigenmode corresponding to a different orientation of the electric dipole mode within the Si spherical particle). The electric and magnetic field magnitude distributions of all degenerate modes are depicted in Figs. S10-S13.The fact that the magnetic field vanishes in the ENZ host can also be more clearly appreciated in those figures. The resonance frequencies and quality factors of these modes have been numerically computed and are depicted in Fig. 3B. Despite the use of realistic losses of SiC ($\varepsilon'' \approx 0.1$), the numerical computation of the resonance frequencies reveals that all degenerate modes in all four cavities deviate less than a 0.3% from the SiC plasma frequency. Furthermore, the fact that degenerate modes exhibit different quality factors allow us to envision the design of a new class of resonators, in which the fields excited by quantum emitters immersed within them exhibit a different decay rate as a function of their polarization, while maintaining the same resonance frequency.

To finalize, there is at least a third set of modes present in ENZ media that are invariant under certain (but not completely arbitrary) geometrical transformations. These modes correspond to the cases where both electric and magnetic fields are neither constant nor zero in the ENZ region. Note that, even if not constant, the magnetic field must always be irrotational $\nabla \times H = -i\omega\varepsilon_0\varepsilon E \approx 0$, and thus it features a quasi-static spatial distribution. However, in this case, the electric field cannot be curl free, $\nabla \times E = i\omega\mu_0 H$, and it indeed takes the form of a solenoidal field forming closed loops in the cavity. The geometry-invariant properties of this set of modes arise from the fact that the modes can concentrate the fields on the vicinity of the internal dielectric particle, resulting in a negligible field at the outer boundaries of the cavity, which naturally satisfies the $\hat{\mathbf{n}} \times E = 0$ boundary condition. In essence, the ENZ properties of the medium ensure that the propagation constant vanishes, $k = \omega\sqrt{\mu_0\varepsilon} \approx 0$, and, hence, the fields cannot propagate through the ENZ host towards the external surface of the cavity. Therefore, when the volume of the particle is much smaller than the volume of the cavity – i.e., $(r_i/r_{out})^3 \ll 1$, with $r_{out}$ being the external radius of a sphere circumscribing the cavity – the field on the surface of the cavity is negligible, and the eigenfrequency becomes approximately independent of the volume and shape of the resonator. Specifically, and as shown in the supplementary online text, section 2.3, the characteristic equation determining the eigenfrequencies of this set of modes can be asymptotically written as follows: $\hat{J}_n\left(\sqrt{\varepsilon_i}\,\frac{\omega}{c}\,r_i\right) = -\left(\sqrt{\varepsilon_i}\,\frac{\omega}{c}\,r_i\right)\hat{J}'_n\left(\sqrt{\varepsilon_i}\,\frac{\omega}{c}\,r_i\right)/n$. Fig. 4 gathers a set of examples illustrating how, as the volume of the cavity increases, the eigenfrequencies converge towards the value prescribed by the characteristic equation. It also exemplifies that once the cavity is sufficiently large, the eigenfrequency becomes independent of the shape of the external surface of the cavity. The examples have been chosen in order to illustrate the impact of progressively increasing the size, and the geometry of the cavities is detailed in Figs. S14-S15. All cavities are composed of a SiC host containing a Si particle, and the properties of the internal particle ($r_i = 1.507$ μm), have been selected so that the eigenfrequency satisfying the characteristic equation for $n = 1$ equals the SiC plasma frequency.

In summary, our theoretical study demonstrates that, in the context of ENZ and zero index metamaterials, there are multiple solutions to the wave equation whose eigenfrequency is invariant under geometrical transformations. It was demonstrated that these solutions enable the existence of geometry-invariant resonant cavities and, hence, they inspire new design philosophies in which the geometry of a device is not determined by, and locked to, its frequency of operation. In this manner, they could also give raise to the next generation of deformable resonant devices. Among other applications, the proposed resonators appear to be particularly well suited for cavity quantum electrodynamics. For instance, the proposed cavities can be locked with their resonances overlapping the atomic transitions of quantum emitters embedded within it, while different aspects of the emitter-cavity interaction can be dynamically tuned by means of deforming the cavity.

This work is supported in part by the US Air Force Office of Scientific Research (AFOSR)

Multidisciplinary University Research Initiative (MURI) on Quantum Metaphotonics &

Metamaterials, Award No. FA9550-12-1-0488.  I. Liberal acknowledges financial support from a FPI scholarship from the Public University of Navarre (UPNA).

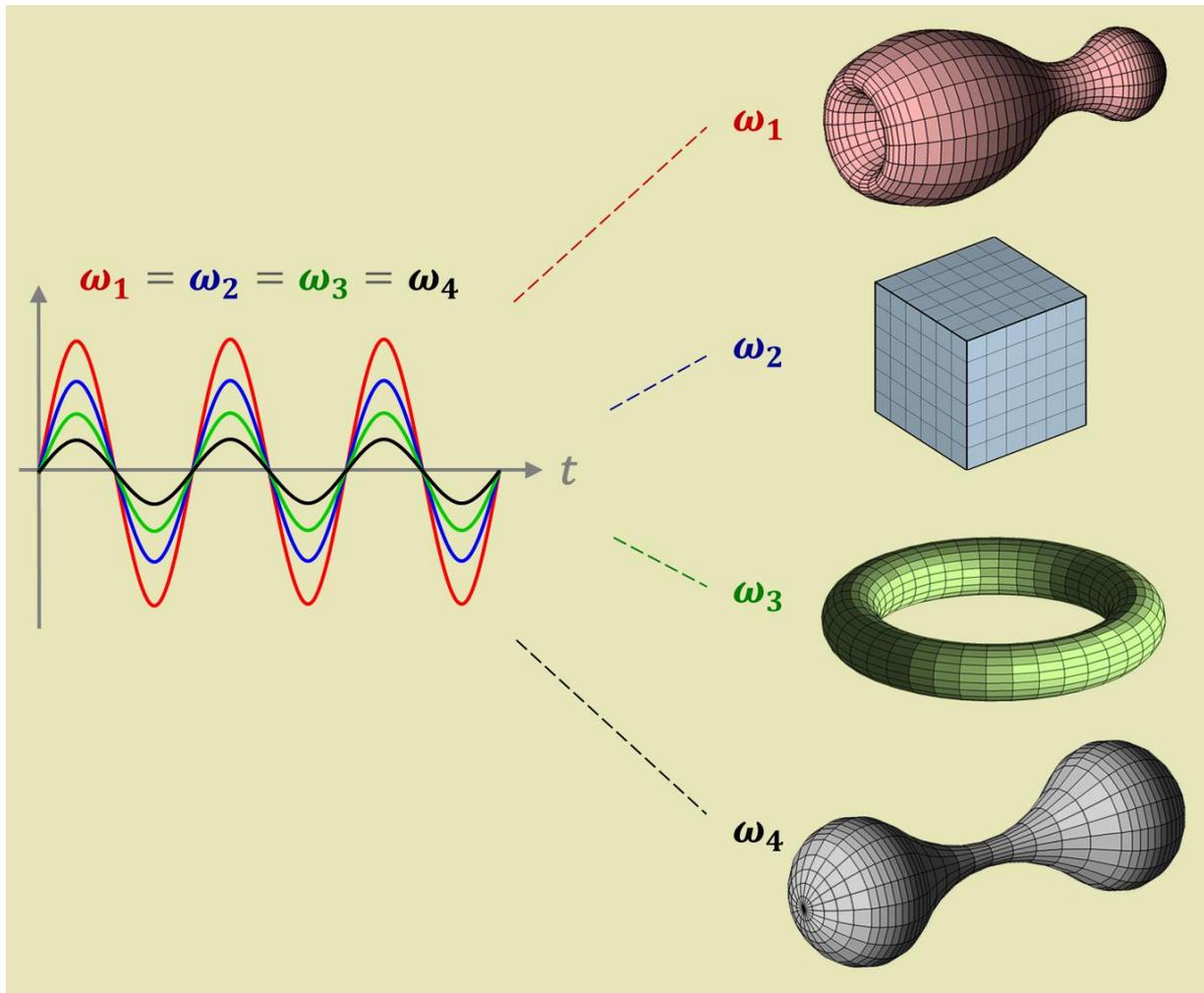

**Fig. 1. Geometry-invariant resonant cavities.** Conceptual sketch of resonant cavities that, despite their very distinct geometry (shape, size, topology…), support an eigenmode at the same resonance frequency.

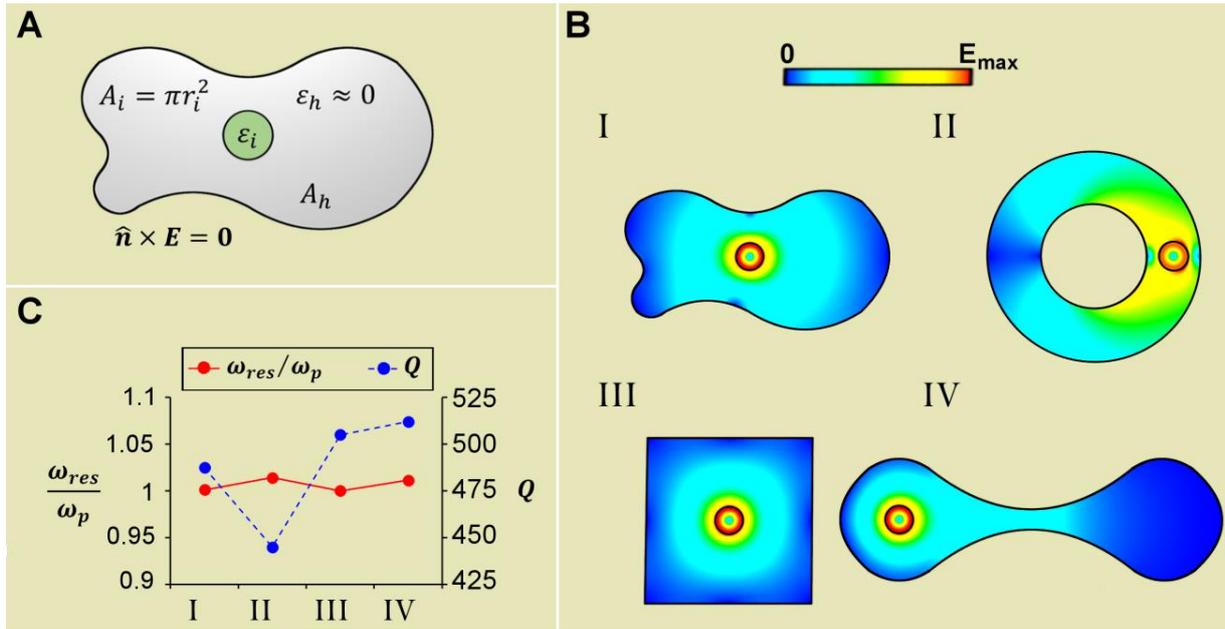

**Fig. 2. Two-dimensional (2D) cavities with eigenfrequencies invariant under equi-areal geometry deformation.** (A) Sketch of a generic 2D cavity composed of an epsilon-near-zero (ENZ) host of cross-sectional area $A_h$ containing an infinitely long circular cylinder of a dielectric (nonmagnetic) material of area $A_i = \pi r_i^2$, and relative permittivity $\varepsilon_i$. The entire cavity is bounded by a perfectly electric conducting (PEC) wall on which the tangential component of $\boldsymbol{E}$ field must vanish. (B): Colormaps of the electric field magnitude distributions of resonant eigenmodes, obtained using numerical simulation, in four cavities consisting of a Si ($\varepsilon_i = 11.7$) cylinder of radius $r_i = 1.165$ µm, immersed in a SiC host of different shapes but equal area $A_i + A_h = 49\pi$ µm$^2$. (C) Linear graph portraying the resonance frequency (normalized to the SiC plasma frequency) and the quality factor, $Q$, as a function of cavity number. Here the imaginary part of relative permittivity of SiC is assumed to be 0.1 at the SiC plasma frequency. In the supplementary materials, we show how these quantities vary with different level of loss (as represented by the different value of the imaginary part of permittivity of SiC).

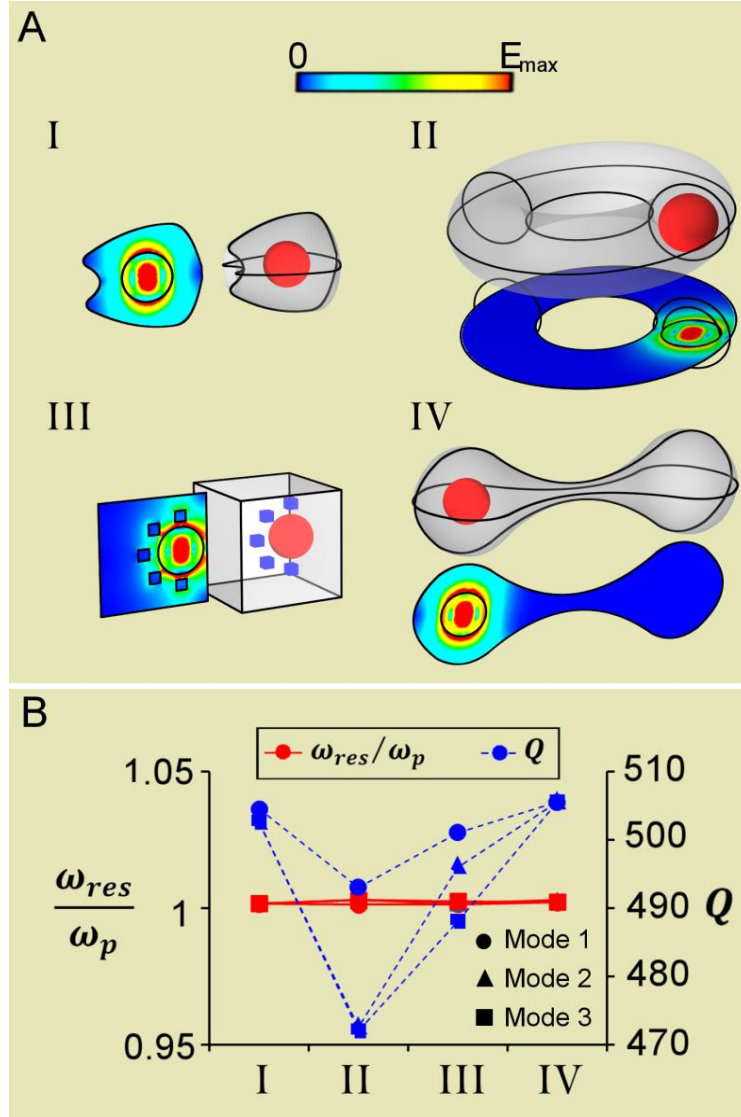

**Fig. 3. Three-dimensional (3D) cavities supporting spatially electrostatic modes with the same resonance eigenfrequency.** (A) Four 3D cavities with different geometries supporting the same resonance eigenfrequency. Each cavity consists of a Si sphere ($\varepsilon_i = 11.7$) (shown as a red sphere) of radius $r_i = 2.155$ μm immersed in a SiC host (shown as grey background), bounded by a PEC wall. Cavity III also contains several additional cubic dielectric particles (shown in blue) with permittivity $\varepsilon_p = 2$, and side $l_p = 1$ μm inserted in the ENZ host. Next to each we show the colormaps of the electric field magnitude distributions obtained using numerical simulation of one of the three supported degenerate eigenmodes (The other eigenmodes can be found in Figs. S10-S13). (B) Linear graph portraying the resonance frequency (normalized to the SiC plasma frequency) and quality factors for these four cavities, demonstrating that the resonance eigenfrequencies are the same, while the quality factors of these eigenmodes are different. Here the imaginary part of relative permittivity of SiC is assumed to be 0.1 at the SiC plasma frequency.

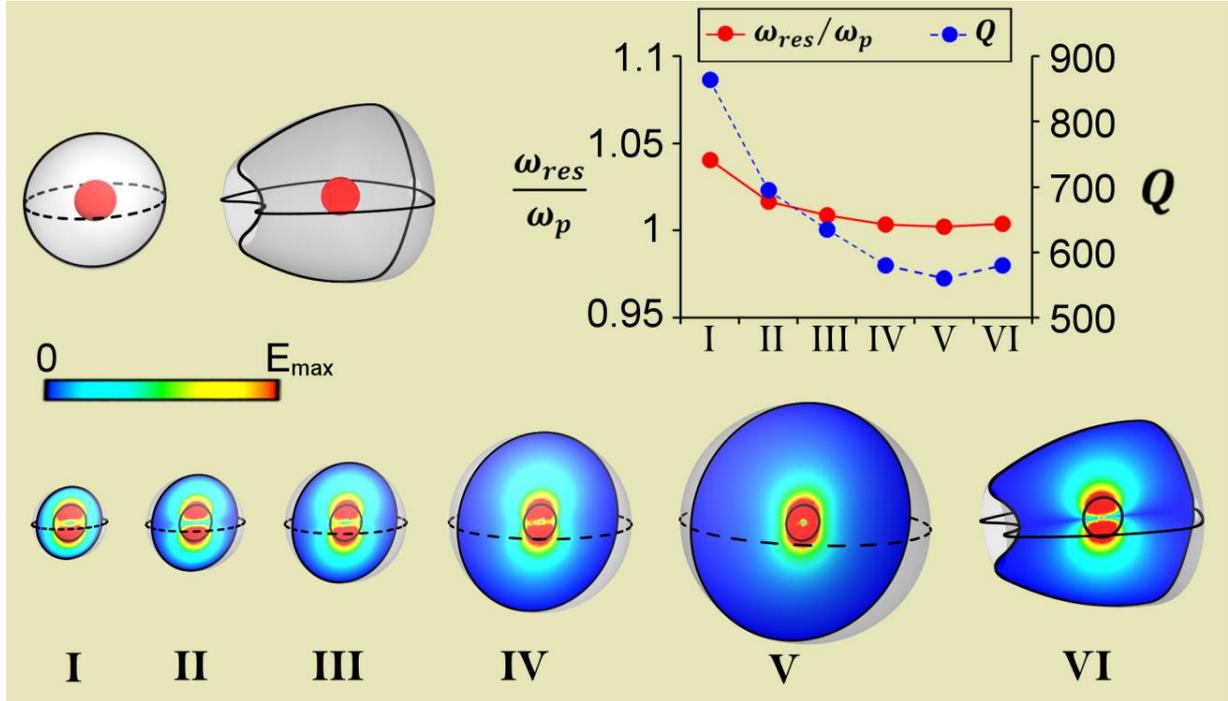

**Fig. 4. Three-dimensional (3D) cavities supporting surface avoiding modes.** Top left panel: Sketches of two cavities consisting of a Si sphere (shown as a red sphere) with ($\varepsilon_i = 11.7$) and radius $r_i = 1.507$ µm immersed in a SiC host (shown as a grey background). Bottom panel: Colormaps of the electric field magnitude distributions of the surface-avoiding eigenmodes, obtained using numerical simulations, in six different cavities (five of which are obtained from the left cavity but with sizes of the ENZ host and the six is the second cavity, shown in the top left panel). Top right panel: linear graph portraying the resonance frequency (normalized to the SiC plasma frequency) and quality factor for six cavities, showing small variation in the resonance frequency, while having different quality factors. Here the imaginary part of relative permittivity of SiC is assumed to be 0.1 at the SiC plasma frequency.

Supplementary Material for:

# Geometry-Invariant Resonant Cavities

**Authors:** I. Liberal[1,2], A. M. Mahmoud[2] and N. Engheta[2*]

**Affiliations:**

[1]Electrical and Electronic Engineering Department, Universidad Pública de Navarra, Pamplona E31006, Spain.

[2]Department of Electrical and Systems Engineering, University of Pennsylvania, Philadelphia, Pennsylvania 19104, USA.

*Correspondence to: engheta@ee.upenn.edu

**This PDF file includes:**

1. Materials and Methods

2. Supplementary Text

3. Supplementary Figures

# 1   Materials and Methods

The commercially available full-wave electromagnetic simulator software COMSOL Multiphysics, version 5.0 [S1], was used in order to generate all 2D and 3D numerical simulations presented in the figures of both the main text and the supplementary material. Specifically, we carried out eigenfrequency analyses in which the software makes use of Finite Element Method (FEM) to determine the eigenmodes and eigenfrequencies of the wave equation $\nabla \times \nabla \times \mathbf{E} - (\omega/c)^2 \varepsilon \mathbf{E} = \mathbf{0}$, subject to be Perfect Electric Conductor (PEC) boundary condition, i.e., the tangential electric field vanishes on the external surface of the cavity, $\hat{\mathbf{n}} \times \mathbf{E} = \mathbf{0}$. The solver was requested to search for eigenfrequencies around the frequency where the real part of the permittivity is near zero ($\varepsilon' \simeq 0$), and the eigenmodes with the closest eigenfrequencies to such a frequency were selected. The silicon (Si) particles were modeled with relative permittivity 11.7 [S2]. The ENZ host was modeled as silicon carbide (SiC) in accordance to [S3]. Thus, the plasma frequency, at which the real part of the permittivity vanishes, $\varepsilon'_{SiC} \simeq 0$, takes place at $\omega_p = 2\pi \times 29.08 \; 10^{12}$ rad/s, with losses represented by $\varepsilon''_{SiC}(\omega_p) \simeq 0.1$.

The numerical solver provided the field distributions of the eigemodes and their associated eigenfrequencies. These data were also employed to calculate the quality factor $Q$ as the ratio of the energies stored and dissipated per unit cycle:

$$Q = \frac{W_{\text{stored}}}{W_{\text{loss}}} \tag{1}$$

where $W_{\text{stored}}$ and $W_{\text{loss}}$ where computed via integration of the electric and magnetic field intensities [S4], taking into account the material dispersion of SiC [S3]

$$W_{\text{loss}} = \frac{1}{4} \int_V \varepsilon'' \varepsilon_0 \, |\mathbf{E}|^2 \, dV \tag{2}$$

$$W_{\text{stored}} = \frac{1}{4} \int_V \left\{ \frac{\partial(\omega \varepsilon')}{\partial \omega} \varepsilon_0 \, |\mathbf{E}|^2 + \mu_0 \, |\mathbf{H}|^2 \right\} dV \tag{3}$$

## 2 Supplementary Text

### 2.1 Eigenfrequencies invariant under equi-areal transformations

Here we demonstrate the existence of 2D cavities supporting an eigenmode whose eigenfrequency is invariant under equi-areal transformations. These eigenmodes are reported in Fig. 2 of the main text. To begin with, let us consider a 2D cavity of area $A = A_h + A_i$, composed by the union of a dielectric cylinder of area $A_i = \pi r_i^2$ and relative permittivity $\varepsilon_i$, immersed in a ENZ host ($\varepsilon_h \simeq 0$) of area $A_h$ (see Fig. 2A). Applying Faraday's law on the surface of the ENZ host we get

$$\oint_{\partial A_h} \mathbf{E} \cdot d\mathbf{l} = \oint_{\partial A} \mathbf{E} \cdot d\mathbf{l} - \oint_{\partial A_i} \mathbf{E} \cdot d\mathbf{l} = i\omega\mu_0 \int_{A_h} \mathbf{H} \cdot \widehat{\mathbf{n}} \, dS \tag{4}$$

The boundary condition $\widehat{\mathbf{n}} \times \mathbf{E} = \mathbf{0}$ imposes that the circulation of the electric field on $\partial A$ is zero. In addition, the magnetic field within the ENZ host must be constant, $\mathbf{H} = \widehat{\mathbf{z}} H_z^h$, in order to ensure a finite electric field $\mathbf{E} = i \left( \omega \varepsilon_0 \varepsilon_h \right)^{-1} \nabla \times \mathbf{H} = i \left( \omega \varepsilon_0 \varepsilon_h \right)^{-1} \nabla H_z^h \times \widehat{\mathbf{z}}$. Therefore, the eigenfrequency of the cavity can be found as follows:

$$\omega = \frac{i}{\mu_0} \frac{\oint_{\partial A_i} \mathbf{E} \cdot d\mathbf{l}}{H_z^h A_h} \tag{5}$$

In order to compute the line integral of the electric field on the surface of the cylinder it is convenient to write the magnetic field in the cylinder as a series of cylindrical harmonics

$$H_z^i = \sum_{n=-\infty}^{\infty} i\omega \, C_n \, J_n \left( k_i \rho \right) e^{jn\phi} \tag{6}$$

where the $C_n$ elements are complex constants determining the magnitude of the magnetic field and $k_i = k_0 \sqrt{\varepsilon_i}$ is the propagation constant within the cylinder. $J_n \left( x \right)$ is the Bessel function of the first kind and order $n$. Recall that the magnetic field in the ENZ region is constant. Therefore, in order to satisfy the continuity of the magnetic field in the surface of the cylinder only the $n = 0$ mode, with no azimuthal variation, can be excited. Exceptions occur when the magnetic field in the ENZ host is zero, cases which are considered in the next section. Thus, by imposing the continuity of the tangential fields for a non-zero magnetic field the line integral of the electric field on the surface of the cylinder can be written as follows:

$$\oint_{\partial A_i} \mathbf{E} \cdot d\mathbf{l} = -i H_z^h \eta_i \left( 2\pi r_i \right) \frac{J_0' \left( k_i r_i \right)}{J_0 \left( k_i r_i \right)} \tag{7}$$

where $\eta_i = \eta_0 / \sqrt{\varepsilon_i}$ is the medium intrinsic impedance within the cylinder. Consequently, the eigenfrequency of the cavity is given by the solution to the following characteristic equation:

$$\omega = \frac{c}{A_h} \frac{2\pi r_i}{\sqrt{\varepsilon_i}} \frac{J_0' \left( k_i r_i \right)}{J_0 \left( k_i r_i \right)} \tag{8}$$

It is apparent from (8) that the eigenfrequency of the cavity is completely determined by the properties of the cylinder and the area of the ENZ host. Therefore, the eigenfrequency is independent of the shape of the external surface of the main cavity as long as the area of the cavity is kept constant. This includes changes in the topology of the cavity such as including holes. The only requirement is that the overall area of the ENZ host is not altered.

## 2.2 Spatially "electrostatic" modes:

Here we demonstrate that epsilon-near-zero ENZ media support $\exp(-i\omega t)$ time-varying spatially electrostatic modes (i.e., $\mathbf{H} = \mathbf{0} \to \nabla \times \mathbf{E} = \mathbf{0}$). These eigenmodes are studied in Fig. 3 of the main text. To begin with, we can inspect time-harmonic Maxwell curl equations in a host medium of permittivity $\varepsilon_h$:

$$\nabla \times \mathbf{H} = -i\omega\varepsilon_0\varepsilon_h\mathbf{E} \tag{9}$$

$$\nabla \times \mathbf{E} = i\omega\mu_0\mathbf{H} \tag{10}$$

It is clear from (9) that, when $\varepsilon_h = 0$, the medium can support non-zero time-varying electric fields even when the magnetic field is zero $\mathbf{H} = \mathbf{0}$. Naturally, it lies from (10) that the electric field must be irrotational $\nabla \times \mathbf{E} = \mathbf{0}$.

The existence of $\exp(-i\omega t)$ time-varying spatially electrostatic modes can also be derived as a limiting case of the usual solution of time-harmonic Maxwell curl equations. Without loss of generality, the electric and magnetic fields can be written as the sum of transversal magnetic ($TM$) and transversal electric ($TE$) fields, i.e., $\mathbf{E} = \mathbf{E}^{TM} + \mathbf{E}^{TE}$ and $\mathbf{H} = \mathbf{H}^{TM} + \mathbf{H}^{TE}$. In addition, each term can be written by using a multipolar decomposition in terms of Tesseral harmonics: [S5]

$$\mathbf{E}^{TM} = -i\sum_{\{q\}}\left\{ n(n+1)\frac{A_{nml}^{TM}\widehat{J}_n(kr) + B_{nml}^{TM}\widehat{Y}_n(kr)}{(kr)^2}\mathbf{T}_{nml}(\widehat{\mathbf{r}}) + \frac{A_{nml}^{TM}\widehat{J}_n'(kr) + B_{nml}^{TM}\widehat{Y}_n'(kr)}{kr}\boldsymbol{\psi}_{nml}(\widehat{\mathbf{r}}) \right\} \tag{11}$$

$$\mathbf{H}^{TM} = \frac{1}{\eta}\sum_{\{q\}}\frac{A_{nml}^{TM}\widehat{J}_n(kr) + B_{nml}^{TM}\widehat{Y}_n(kr)}{kr}\boldsymbol{\varphi}_{nml}(\widehat{\mathbf{r}}) \tag{12}$$

$$\mathbf{E}^{TE} = \sum_{\{q\}}\frac{A_{nml}^{TE}\widehat{J}_n(kr) + B_{nml}^{TE}\widehat{Y}_n(kr)}{kr}\boldsymbol{\varphi}_{nml}(\widehat{\mathbf{r}}) \tag{13}$$

$$\mathbf{H}^{TE} = \frac{i}{\eta}\sum_{\{q\}}\left\{ n(n+1)\frac{A_{nml}^{TE}\widehat{J}_n(kr) + B_{nml}^{TE}\widehat{Y}_n(kr)}{(kr)^2}\mathbf{T}_{nml}(\widehat{\mathbf{r}}) + \frac{A_{nml}^{TE}\widehat{J}_n'(kr) + B_{nml}^{TE}\widehat{Y}_n'(kr)}{kr}\boldsymbol{\varphi}_{nml}(\widehat{\mathbf{r}}) \right\} \tag{14}$$

where $\{q\} = \{n, m, l\}$ is a multi-index defined so that the sum runs over all spherical multipoles:

$$\sum_{\{q\}} = \sum_{n=1}^{\infty}\sum_{m=0}^{n}\sum_{l=e,o} \tag{15}$$

In such a decomposition, the radial dependency of the field is described via the Schelkunoff form of the spherical Bessel functions, $\widehat{J}_n(kr)$ and $\widehat{Y}_n(kr)$, where, for example, $\widehat{J}_n(x) = \sqrt{\pi x/2}\,J_{n+1/2}(x)$ with $J_n(x)$ being the Bessel function of the first kind and order $n$ [S5]. On the other hand, the angular dependency is described by Tesseral harmonics $\mathbf{T}_{nml}(\widehat{\mathbf{r}})$ and linear combinations of its derivatives, $\boldsymbol{\psi}_{nml}(\widehat{\mathbf{r}})$ and $\boldsymbol{\varphi}_{nml}(\widehat{\mathbf{r}})$, given by:

$$\mathbf{T}_{nm\binom{e}{o}}(\widehat{\mathbf{r}}) = \widehat{\mathbf{r}}\,P_n^m(\cos\theta)\binom{\cos m\phi}{\sin m\phi} \tag{16}$$

$$\boldsymbol{\psi}_{nml}(\widehat{\mathbf{r}}) = \widehat{\boldsymbol{\theta}}\,\partial_\theta T_{nml}(\widehat{\mathbf{r}}) + \widehat{\boldsymbol{\phi}}\frac{\partial_\phi T_{nml}(\widehat{\mathbf{r}})}{\sin\theta} \tag{17}$$

$$\boldsymbol{\varphi}_{nml}(\widehat{\mathbf{r}}) = \boldsymbol{\psi}_{nml}(\widehat{\mathbf{r}}) \times \widehat{\mathbf{r}} \tag{18}$$

When the host medium is ENZ ($\varepsilon \to 0$, so $\eta \to \infty$ and $k \to 0$), the TM fields (19)-(20) reduce to

$$\mathbf{E}^{TM} = -i \sum_{\{q\}} \left\{ n\left(n+1\right) \left[ \frac{1}{(2n+1)!!} \left(k^{n-1} A_{nml}^{TM}\right) r^{n-1} - (2n-1)!! \left(\frac{B_{nml}^{TM}}{k^{n+2}}\right) \frac{1}{r^{n+2}} \right] \mathbf{T}_{nml}\left(\widehat{\mathbf{r}}\right) \right.$$

$$\left. + \left[ \frac{n+1}{(2n+1)!!} \left(k^{n-1} A_{nml}^{TM}\right) r^{n-1} + n\left(2n-1\right)!! \left(\frac{B_{nml}^{TM}}{k^{n+2}}\right) \frac{1}{r^{n+2}} \right] \boldsymbol{\psi}_{nml}\left(\widehat{\mathbf{r}}\right) \right\} \tag{19}$$

$$\mathbf{H}^{TM} = \frac{1}{\omega\mu_0} \sum_{\{q\}} \left[ \frac{1}{(2n+1)!!} \left(k^{n+1} A_{nml}^{TM}\right) r^{n} - (2n-1)!! \left(\frac{B_{nml}^{TM}}{k^{n}}\right) \frac{1}{r^{n-1}} \right] \boldsymbol{\varphi}_{nml}\left(\widehat{\mathbf{r}}\right) \tag{20}$$

It is apparent from (19) that in order to $\mathbf{E}^{TM}$ to be finite but non-zero it is required $A_{nml}^{TM} \propto k^{-n+1}$ and $B_{nml}^{TM} \propto k^{n+2}$. However, it is also clear from (12) that this implies a zero magnetic field $\mathbf{H}^{TM} = \mathbf{0}$. In this manner, the multipolar decomposition ratifies the existence spatially electrostatic fields in ENZ media. In addition, this decomposition gives some hints on how to excite these fields. In general, sources of $TM$ multipoles and their linear combinations generate time-varying spatially electrostatic fields in ENZ.

On the other hand, the TE fields (21)-(22) in ENZ media can be written as follows:

$$\mathbf{E}^{TE} = \sum_{\{q\}} \left[ \frac{1}{(2n+1)!!} \left(k^{n} A_{nml}^{TE}\right) r^{n} - (2n-1)!! \left(\frac{B_{nml}^{TE}}{k^{n+1}}\right) \frac{1}{r^{n+1}} \right] \boldsymbol{\varphi}_{nml}\left(\widehat{\mathbf{r}}\right) \tag{21}$$

$$\mathbf{H}^{TE} = \frac{i}{\omega\mu_0} \sum_{\{q\}} \left\{ n\left(n+1\right) \left[ \frac{1}{(2n+1)!!} \left(k^{n} A_{nml}^{TE}\right) r^{n-1} - (2n-1)!! \left(\frac{B_{nml}^{TE}}{k^{n+1}}\right) \frac{1}{r^{n+2}} \right] \mathbf{T}_{nml}\left(\widehat{\mathbf{r}}\right) \right.$$

$$\left. \left[ \frac{n+1}{(2n+1)!!} \left(k^{n} A_{nml}^{TE}\right) r^{n-1} + n\left(2n-1\right)!! \left(\frac{B_{nml}^{TE}}{k^{n+1}}\right) \frac{1}{r^{n+2}} \right] \boldsymbol{\psi}_{nml}\left(\widehat{\mathbf{r}}\right) \right\} \tag{22}$$

In this case, in order to obtain a finite but non-zero elecric field $\mathbf{E}^{TE}$ requires $A_{nml}^{TE} \propto k^{-n}$ and $B_{nml}^{TE} \propto k^{n+1}$. However, by contrast with the TM modes, this does not implies a zero mangetic field $\mathbf{H}^{TE}$.

## 2.3 Spherical cavities

Specific examples help to understand the behavior of some of the modes present in cavities of arbitrary geometry. To this end, let us consider a cavity consisting of two concentric spheres, with internal and external radii of $a$ and $b$, respectively, and with internal and external permittivities of $\varepsilon_1$ and $\varepsilon_2$, respectively (see Fig. S16). Due to its spherical symmetry, the modes excited in the cavity correspond to the spherical harmonics (11)-(14). The characteristic equations that determine the eigenfrequencies of each spherical harmonic can be found by imposing the boundary condition $\hat{\mathbf{n}} \times \mathbf{E} = \mathbf{0}$ on the surface of the cavity, and enforcing the continuity of the fields on the surface between the two concentric spheres. This exercise leads to the following characteristic equations:

$$\frac{\widehat{Y}'_n(k_2 b)\widehat{J}'_n(k_2 a) - \widehat{J}'_n(k_2 b)\widehat{Y}'_n(k_2 a)}{\widehat{Y}'_n(k_2 b)\widehat{J}_n(k_2 a) - \widehat{J}'_n(k_2 b)\widehat{Y}_n(k_2 a)} = \frac{\eta_1}{\eta_2}\frac{\widehat{J}'_n(k_1 a)}{\widehat{J}_n(k_1 a)}, \text{ for } TM \text{ modes} \tag{23}$$

$$\frac{\widehat{Y}_n(k_2 b)\widehat{J}_n(k_2 a) - \widehat{J}_n(k_2 b)\widehat{Y}_n(k_2 a)}{\widehat{Y}_n(k_2 b)\widehat{J}'_n(k_2 a) - \widehat{J}_n(k_2 b)\widehat{Y}'_n(k_2 a)} = \frac{\eta_1}{\eta_2}\frac{\widehat{J}_n(k_1 a)}{\widehat{J}'_n(k_1 a)}, \text{ for } TE \text{ modes} \tag{24}$$

where $\eta_1$, $k_1$ and $\eta_2$, $k_2$ are the medium impedance and propagation constant of the internal and external regions, respectively. When the outer layer of the cavity is filled with a ENZ material ($\varepsilon_2 \to 0$), the characteristic equations can be asymptotically written as follows

$$\frac{\left(\frac{a}{b}\right)^{2n+1} - 1}{\frac{1}{n+1}\left(\frac{a}{b}\right)^{2n+1} + \frac{1}{n}} = \frac{\varepsilon_2}{\varepsilon_1}\,(k_1 a)\,\frac{\widehat{J}'_n(k_1 a)}{\widehat{J}_n(k_1 a)}, \text{ for } TM \text{ modes} \tag{25}$$

$$\frac{\left(\frac{a}{b}\right)^{2n+1} - 1}{(n+1)\left(\frac{a}{b}\right)^{2n+1} + n} = \frac{1}{k_1 a}\,\frac{\widehat{J}_n(k_1 a)}{\widehat{J}'_n(k_1 a)}, \text{ for } TE \text{ modes} \tag{26}$$

Note that the r.h.s. of (25) goes to zero since $\varepsilon_2 \to 0$, and hence, there is no solution for the $TM$ modes except at $\widehat{J}_n(k_1 a) = 0$. This condition corresponds to a zero tangential magnetic field, which enables the existence of spatially electrostatic fields characterizing $TM$ modes in the ENZ region, while at the same time satisfying the boundary conditions on the surface of the cavity.

Interestingly, when the volume of the internal sphere is much smaller than the volume of the whole cavity, $(a/b)^{2n+1} \ll 1$, the characteristic equations (25)-(26) can be approximated as follows

$$-n = \frac{\varepsilon_2}{\varepsilon_1}\,(k_1 a)\,\frac{\widehat{J}'_n(k_1 a)}{\widehat{J}_n(k_1 a)}, \text{ for } TM \text{ modes} \tag{27}$$

$$-\frac{1}{n} = \frac{1}{k_1 a}\,\frac{\widehat{J}_n(k_1 a)}{\widehat{J}'_n(k_1 a)}, \text{ for } TE \text{ modes} \tag{28}$$

Note that in this limit the characteristic equation becomes independent of the external surface even for the $TE$ modes.

## 2.4 Perturbational Techniques

Here we use perturbational techniques (see, e.g., [S5]) to ratify further that the eigenfrequencies of spatially electrostatic eigenmodes are preserved under geometrical transformations. To this end, let us consider a cavity of volume $V$ enclosed within a surface $S$ subject to the boundary condition $\widehat{\mathbf{n}} \times \mathbf{E} = \mathbf{0}$ (see Fig. S17). We assume that this cavity supports an eigenmode with fields $\mathbf{E}_0, \mathbf{H}_0$ at the eigenfrequency $\omega_0$. The cavity could be, for example, the spherical cavity described in the previous section. Next, we generate a second cavity by modifying the original cavity, leading to a volume $V' = V - \triangle V$ and $S' = S - \triangle S$. In general, this new cavity will support a different eigenmode with fields $\mathbf{E}, \mathbf{H}$ at an also different eigenfrequency $\omega$. Note that the fields of the eigenmode in the first cavity satisfy time-harmonic curl Maxwell equations at frequency $\omega_0$:

$$\nabla \times \mathbf{E}_0 = i\omega_0 \mu_0 \mathbf{H}_0 \tag{29}$$

$$\nabla \times \mathbf{H}_0 = -i\omega_0 \varepsilon_0 \varepsilon \mathbf{E}_0 \tag{30}$$

On the other hand, the fields of the eigenmode in the second cavity satisfy time-harmonic curl Maxwell equations at frequency $\omega$:

$$\nabla \times \mathbf{E} = i\omega \mu_0 \mathbf{H} \tag{31}$$

$$\nabla \times \mathbf{H} = -i\omega \varepsilon_0 \varepsilon \mathbf{E} \tag{32}$$

In this manner, we can then write:

$$\nabla \cdot (\mathbf{H} \times \mathbf{E}_0^*) = \mathbf{H} \cdot \nabla \times \mathbf{E}_0^* - \mathbf{E}_0^* \cdot \nabla \times \mathbf{H} = -i\omega_0 \mu_0 \mathbf{H} \cdot \mathbf{H}_0^* + i\omega \varepsilon_0 \varepsilon \mathbf{E}_0^* \cdot \mathbf{E} \tag{33}$$

$$\nabla \cdot (\mathbf{H}_0^* \times \mathbf{E}) = \mathbf{H}_0^* \cdot \nabla \times \mathbf{E} - \mathbf{E} \cdot \nabla \times \mathbf{H}_0^* = i\omega \mu_0 \mathbf{H}_0^* \cdot \mathbf{H} - i\omega_0 \varepsilon_0 \varepsilon \mathbf{E} \cdot \mathbf{E}_0^* \tag{34}$$

Next, integrating over $V'$, and applying the divergence theorem we get

$$\oint_{S'} (\mathbf{H} \times \mathbf{E}_0^* + \mathbf{H}_0^* \times \mathbf{E}) \cdot \widehat{\mathbf{n}} dS' = i(\omega - \omega_0) \int_{V'} (\mu_0 \mathbf{H}_0^* \cdot \mathbf{H} + \varepsilon_0 \mathbf{E}_0^* \cdot \mathbf{E}) \, dV' \tag{35}$$

However, since $\widehat{\mathbf{n}} \times \mathbf{E} = \mathbf{0}$ on $S'$ we have

$$\oint_{S'} (\mathbf{H} \times \mathbf{E}_0^* + \mathbf{H}_0^* \times \mathbf{E}) \cdot \widehat{\mathbf{n}} dS' = \oint_{S'} (\mathbf{H} \times \mathbf{E}_0^*) \cdot \widehat{\mathbf{n}} dS' \tag{36}$$

Moreover, since $\widehat{\mathbf{n}} \times \mathbf{E}_0 = \mathbf{0}$ on $S$ and $S' = S - \Delta S$ we get

$$\oint_{S'} (\mathbf{H} \times \mathbf{E}_0^*) \cdot \widehat{\mathbf{n}} dS' = -\oint_{\triangle S} (\mathbf{H} \times \mathbf{E}_0^*) \cdot \widehat{\mathbf{n}} dS' \tag{37}$$

Therefore, the shift in the resonance frequency is given by

$$\omega - \omega_0 = \frac{i \oint_{\triangle S} (\mathbf{H} \times \mathbf{E}_0^*) \cdot \widehat{\mathbf{n}} dS'}{\int_{V'} (\mu_0 \mathbf{H}_0^* \cdot \mathbf{H} + \varepsilon_0 \mathbf{E}_0^* \cdot \mathbf{E}) \, dV'} \tag{38}$$

We emphasize that (38) is not an approximation. It is a exact expression of the frequency shift due to the deformation. At the same time, (38) can only be evaluated with a complete knowledge of the eigenmode fields $\mathbf{E}_0$, $\mathbf{H}_0$, $\mathbf{E}$, $\mathbf{H}$. In general, one applies asumptions on the form of those fields to estimate the frequency shift. For example, a common approximation is to replace the perturbed eigenmode fields $\mathbf{E}$, $\mathbf{H}$ by the unperturbed eigenmode fields $\mathbf{E}_0$, $\mathbf{H}_0$, which is a reasonable estimate for small deformations [S5]. In our case, if the deformation takes place within the ENZ host, there is no change in the electrostatic nature on the fields, so that $\mathbf{H} = \mathbf{0}$ in $\triangle S$. Thus, it lies from (38) that when $\mathbf{H} = \mathbf{0}$ in $\triangle S$ and there is no change in the resonance frequency.

**3 Supplementary Figures:**

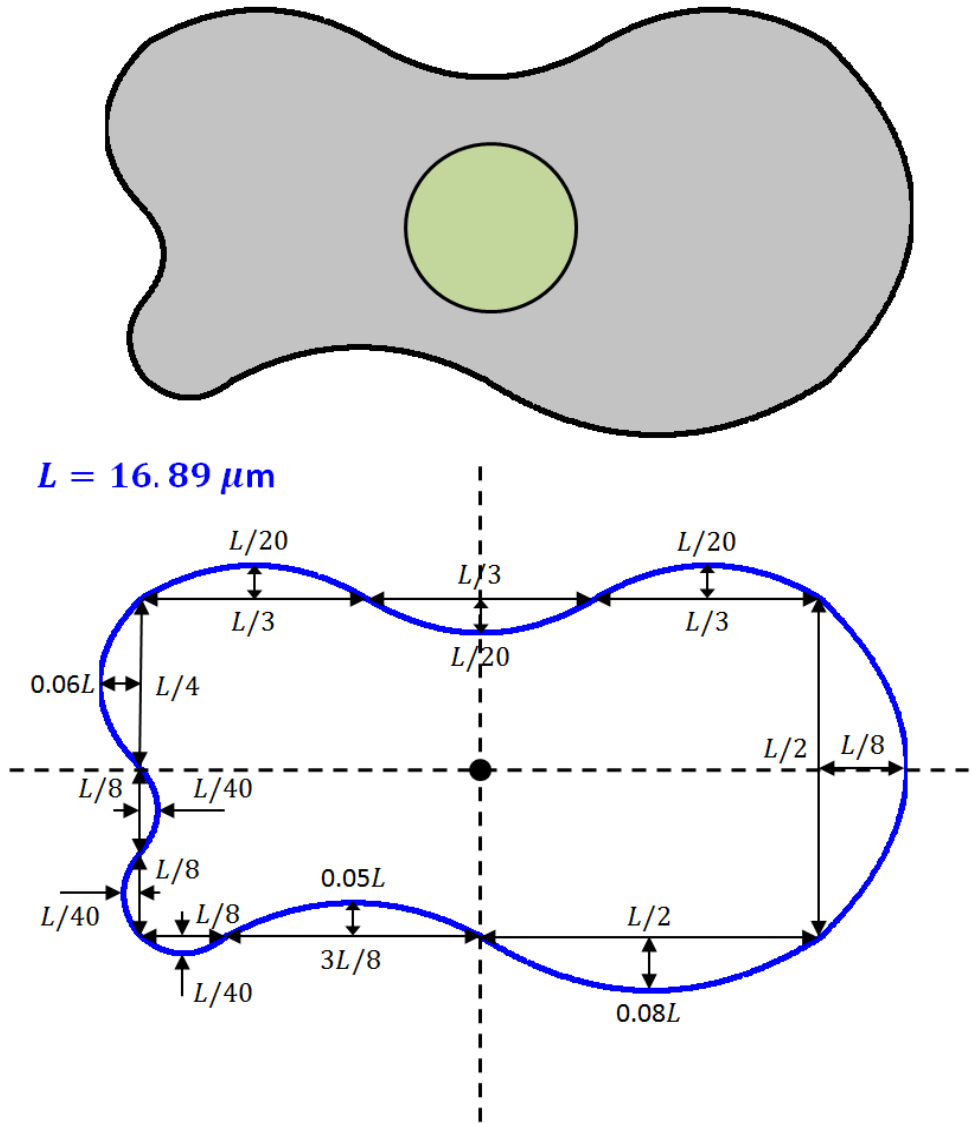

**Fig. S1.** Sketch (top) and dimensions (bottom) of the cavity studied in Fig. 2B-I of the main text. The cavity consists of a Si cylinder ($\varepsilon_i = 11.7$, shown in green) of radius $r_i = 1.165$ μm immersed in a SiC host (shown as grey background). The location of the sphere in the main cavity is indicated by a black dot. The blue curves correspond to the second order polynomials $f(u) = c_0 + c_1 u + c_2 u^2$ that fit to the specified dimensions. The resulting area is $\pi 7^2$ μm.

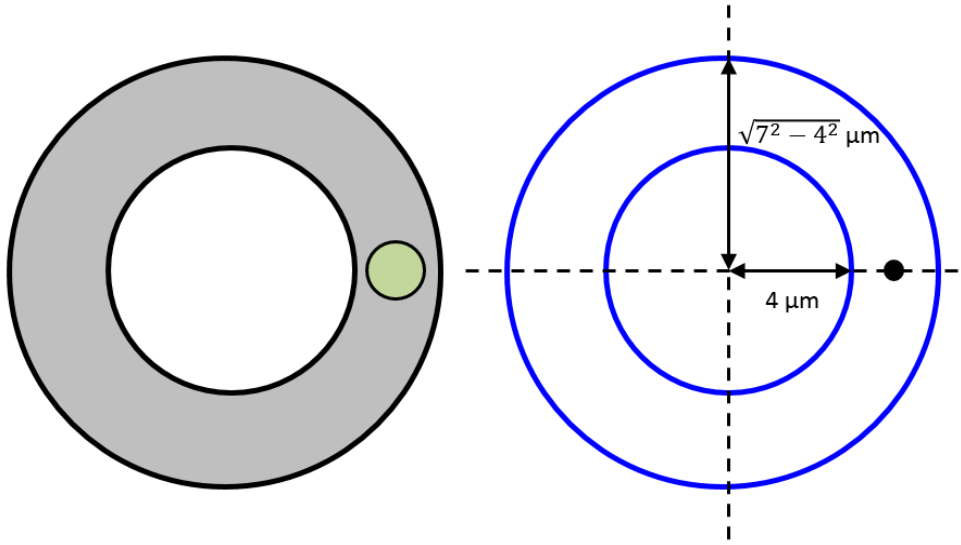

**Fig. S2.** Sketch (left) and dimensions (right) of the cavity studied in Fig. 2B-II of the main text. The cavity consists of a Si cylinder ($\varepsilon_i = 11.7$) (shown in green) of radius $r_i = 1.165$ μm immersed in a SiC host (shown as grey background). The location of the sphere in the main cavity is indicated by a black dot. The area contained between the two blue circles conforming the main cavity also equals $\pi 7^2$ μm.

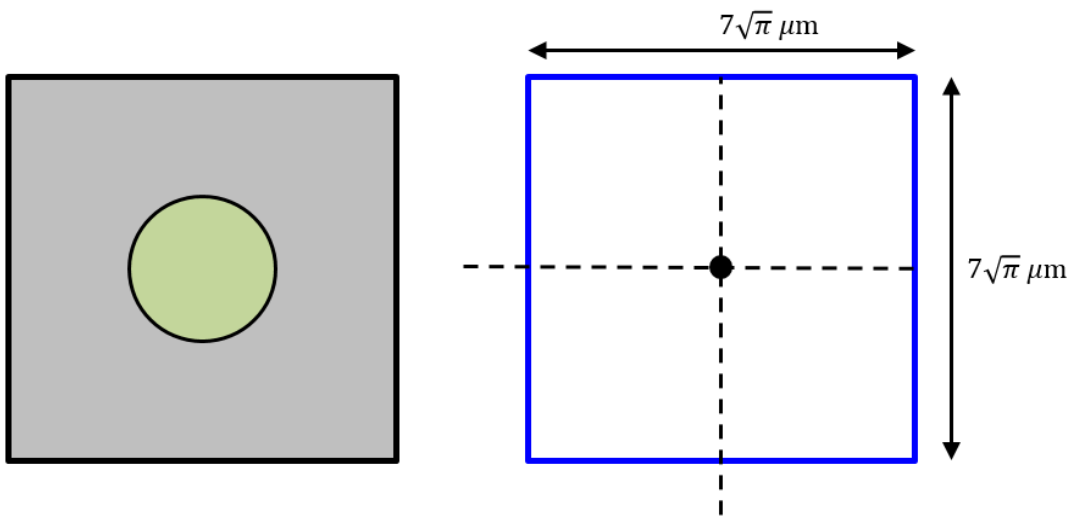

**Fig. S3.** Sketch (left) and dimensions (right) of the cavity studied in Fig. 2B-III of the main text. The cavity consists of a Si cylinder ($\varepsilon_i = 11.7$) (shown in green) of radius $r_i = 1.165$ μm immersed in a SiC host (shown as grey background). The location of the sphere in the main cavity is indicated by a black dot. The area of the main cavity also equals $\pi 7^2$ μm.

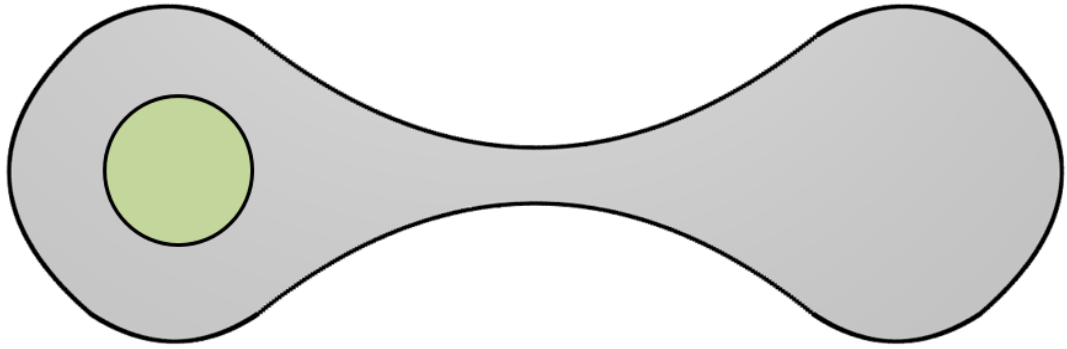

$L = 15.2 \ \mu m$

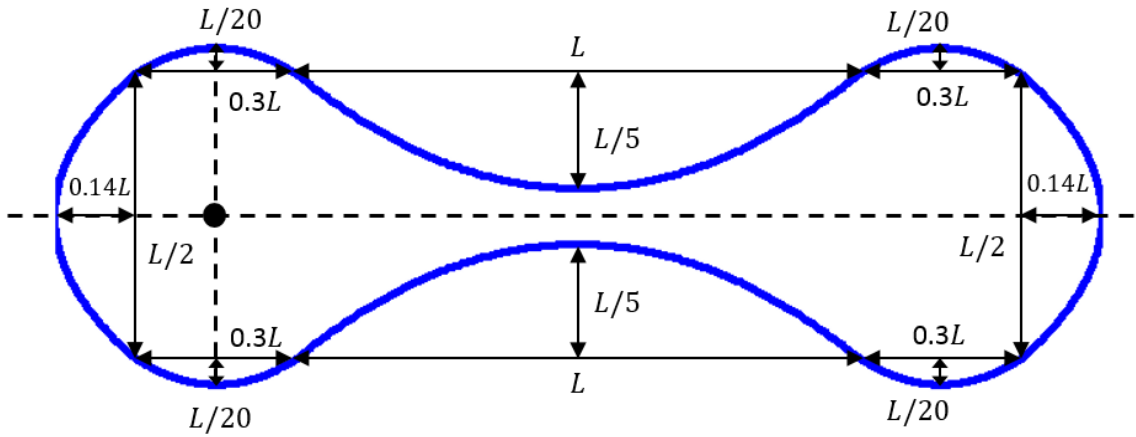

**Fig. S4.** Sketch (top) and dimensions (bottom) of the cavity studied in Fig. 2B-IV of the main text. The cavity consists of a Si cylinder ($\varepsilon_i = 11.7$) (shown in green) of radius $r_i = 1.165$ μm immersed in a SiC host (shown as grey background). The location of the sphere in the main cavity is indicated by a black dot. The blue curves correspond to the second order polynomials $f(u) = c_0 + c_1 u + c_2 u^2$ that fit to the specified dimensions. The resulting area is also $\pi 7^2 \ \mu m$.

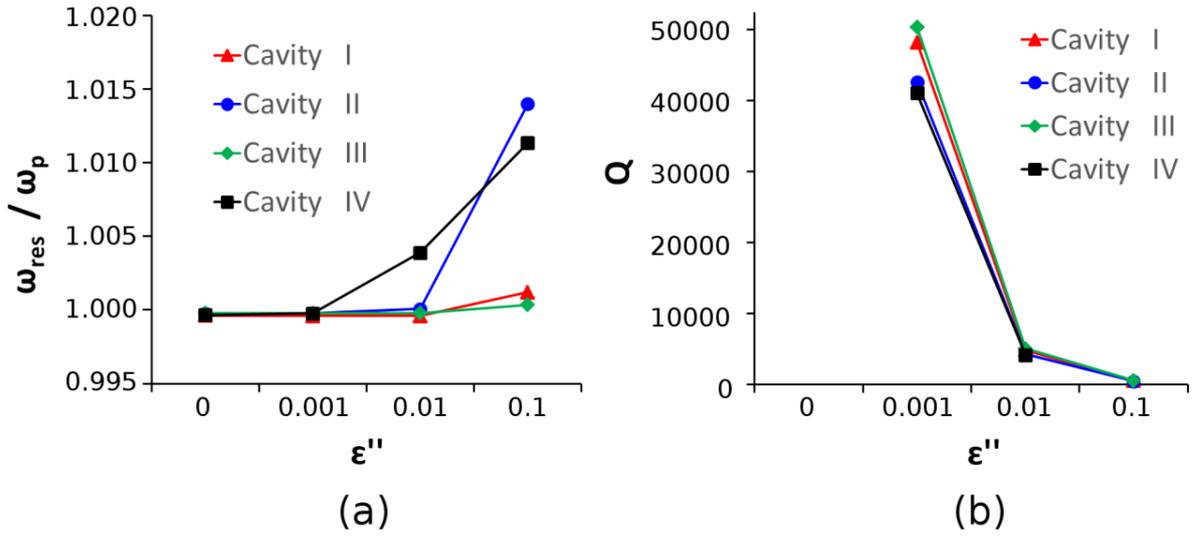

**Fig. S5.** Simulation results for (a) resonance frequency, normalized to the SiC plasma frequency, and (b) quality factor of the cavities represented in Fig. 2C of the main text for different amounts of loss at the SiC plasma frequency.

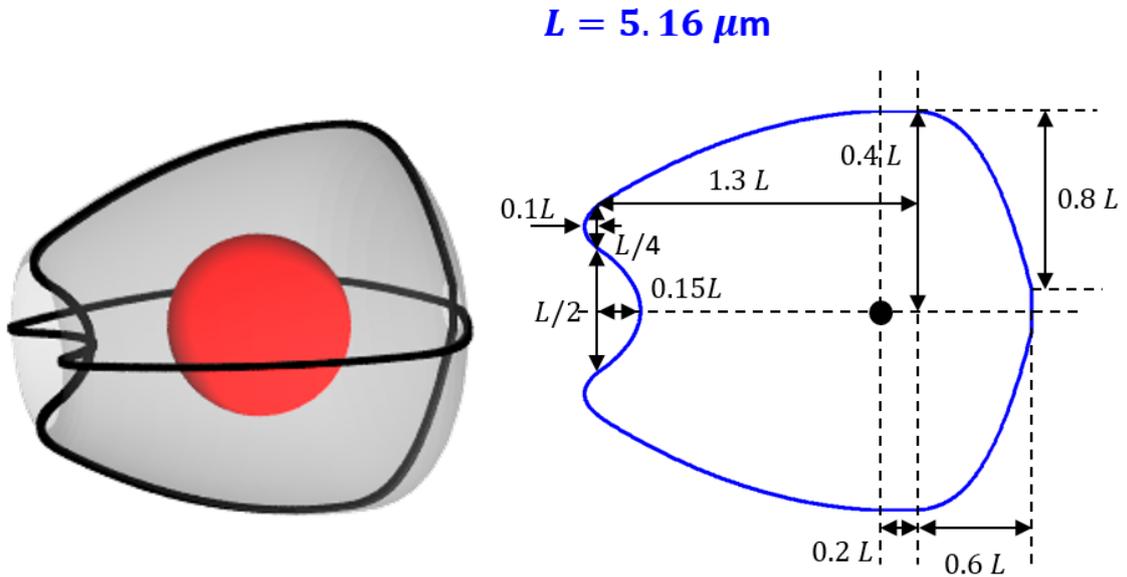

**Fig. S6.** Sketch (left) and dimensions (right) of the cavity studied in Fig. 3A-I of the main text. The cavity consists of a Si sphere ($\varepsilon_i = 11.7$) (shown in red) of radius $r_i = 2.155$ μm immersed in a SiC host (shown as grey background). The location of the sphere in the main cavity is indicated by a black dot.

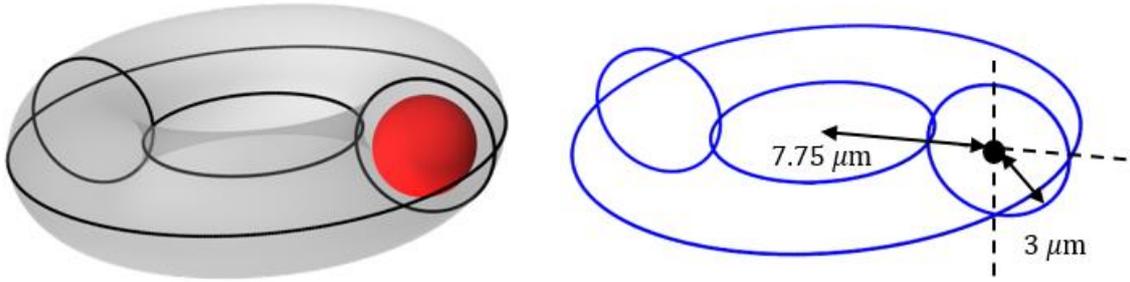

**Fig. S7.** Sketch (left) and dimensions (right) of the cavity studied in Fig. 3A-II of the main text. The cavity consists of a Si sphere ($\varepsilon_i = 11.7$) (shown in red) of radius $r_i = 2.155$ µm immersed in a SiC host (shown as grey background). The location of the sphere in the main cavity is indicated by a black dot.

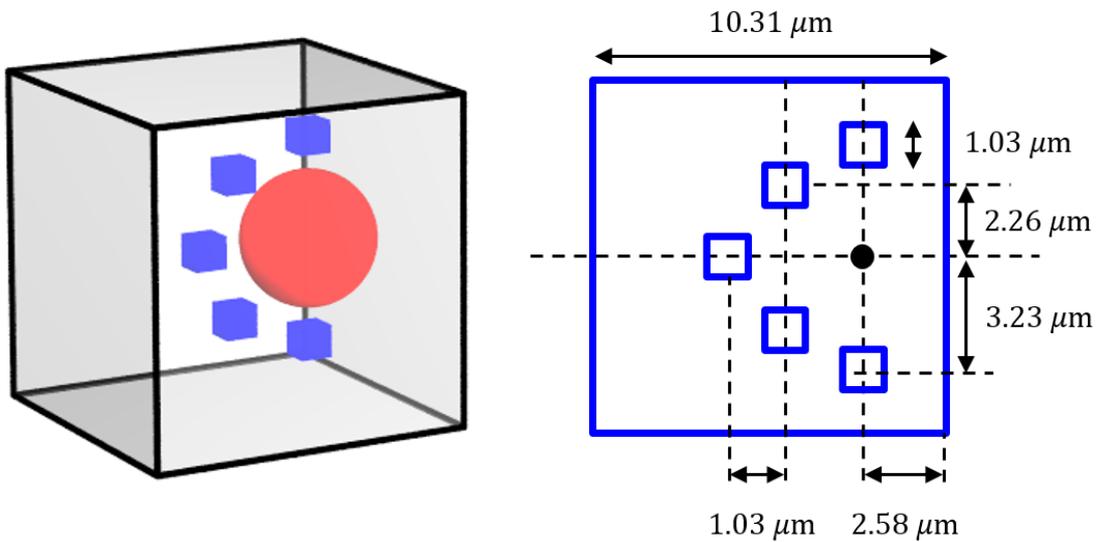

**Fig. S8.** Sketch (left) and dimensions (right) of the cavity studied in Fig. 3A-III of the main text. The cavity consists of a Si sphere ($\varepsilon_i = 11.7$) (shown in red) of radius $r_i = 2.155$ µm immersed in a SiC host (shown as grey background). The cavity also contains several additional cubic dielectric particles (shown in blue) with permittivity $\varepsilon_p = 2$. The location of the sphere in the main cavity is indicated by a black dot.

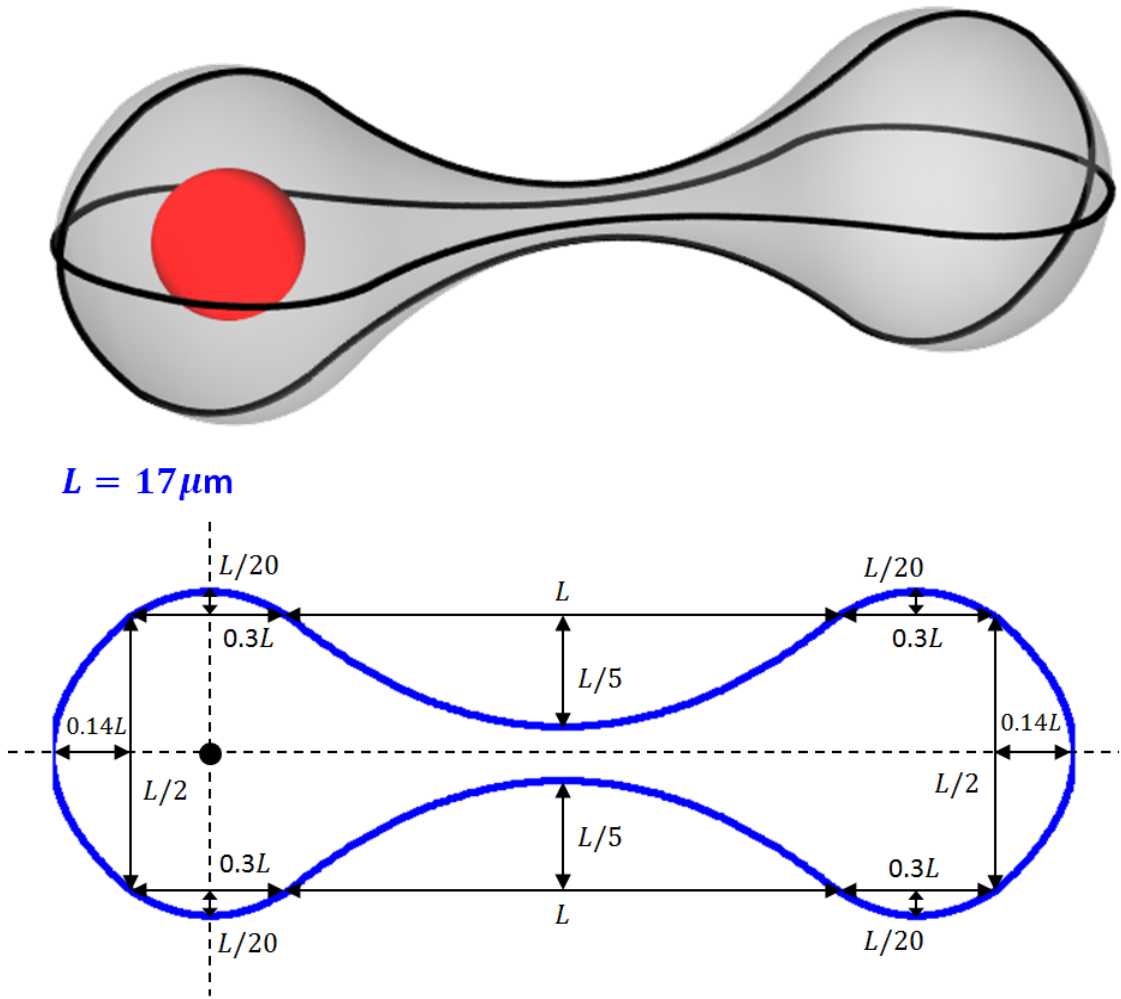

**Fig. S9.** Sketch (left) and dimensions (right) of the cavity studied in Fig. 3A-IV of the main text. The cavity consists of a Si sphere ($\varepsilon_i = 11.7$) (shown in red) of radius $r_i = 2.155$ µm immersed in a SiC host (shown as grey background). The location of the sphere in the main cavity is indicated by a black dot.

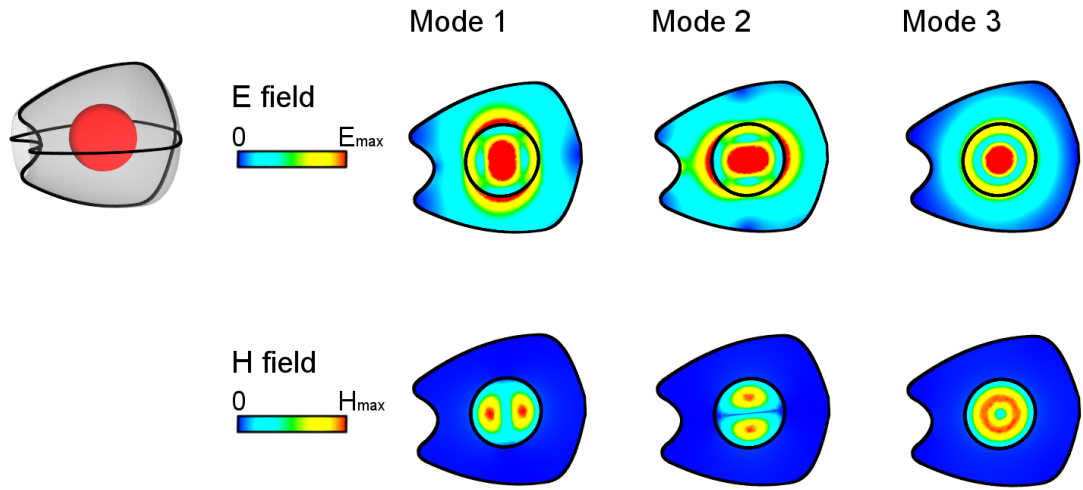

**Fig. S10.** Colormaps of the electric and magnetic field magnitude distributions of the degenerate modes excited in the cavity depicted in Fig. 3A-I of the main text. The eigenfrequencies and quality factors of these eigenmodes are shown in Fig. 3B, cavity I, of the main text.

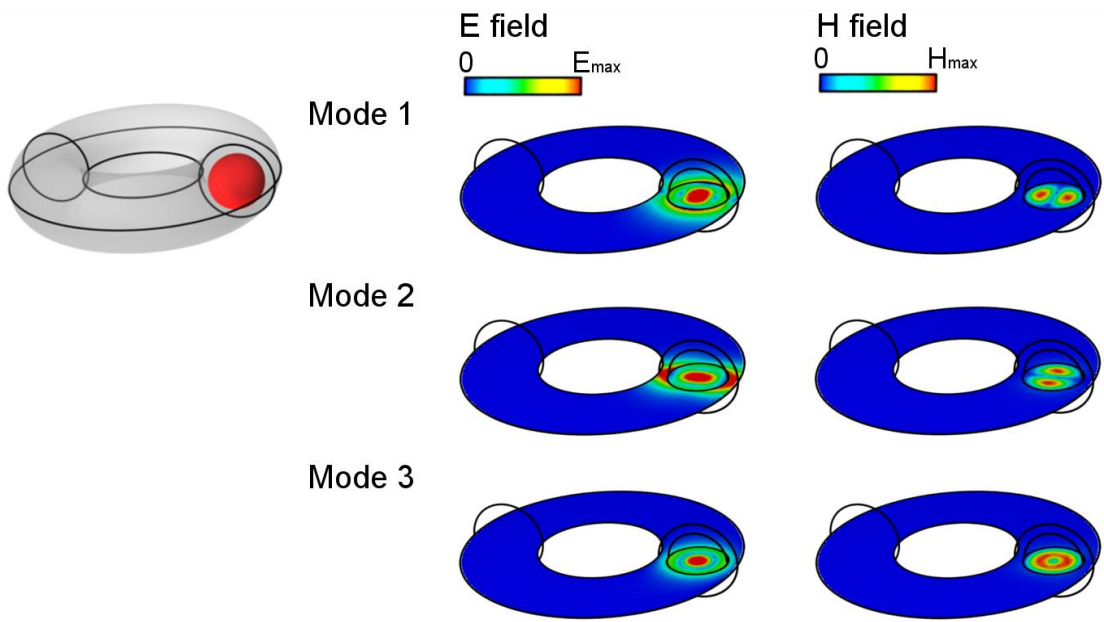

**Fig. S11.** Colormaps of the electric and magnetic field magnitude distributions of the degenerate modes excited in the cavity depicted in Fig. 3A-II of the main text. The eigenfrequencies and quality factors of these eigenmodes are shown in Fig. 3B, cavity II, of the main text.

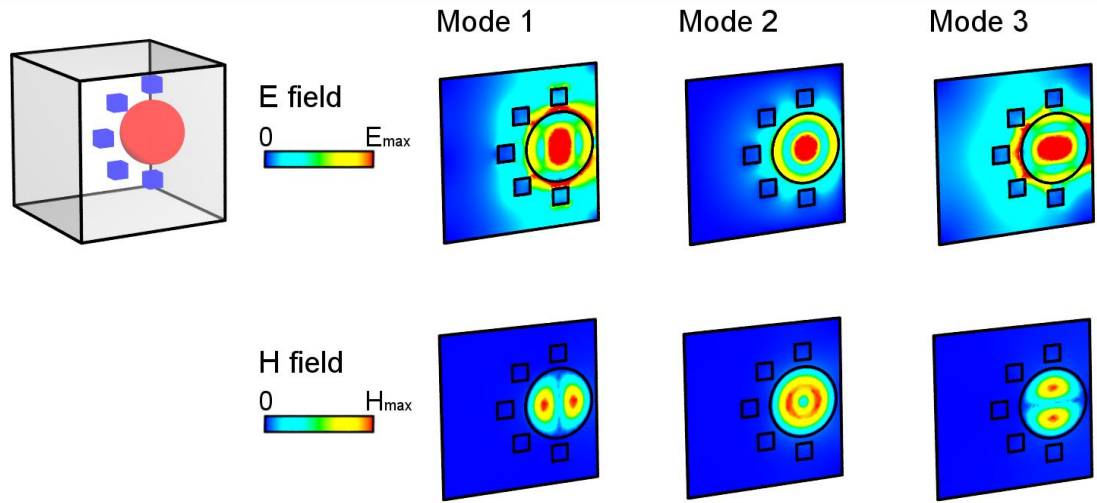

**Fig. S12.** Colormaps of the electric and magnetic field magnitude distributions of the degenerate modes excited in the cavity depicted in Fig. 3A-III, of the main text. The eigenfrequencies and quality factors of these eigenmodes are shown in Fig. 3B, cavity III, of the main text.

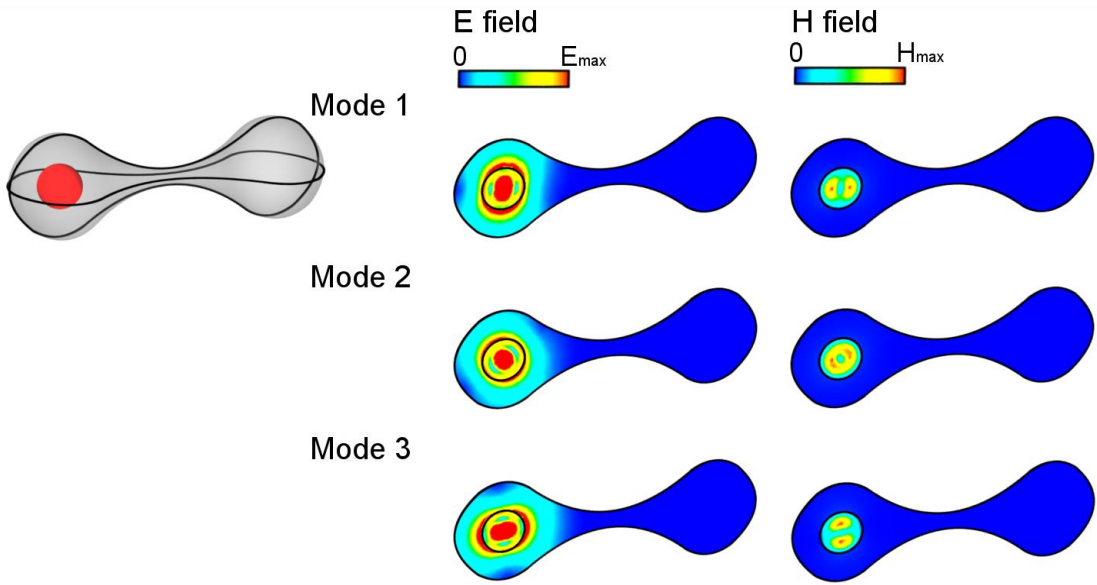

**Fig. S13.** Colormaps of the electric and magnetic field magnitude distributions of the degenerate modes excited in the cavity depicted in Fig. 3A-IV, of the main text. The eigenfrequencies and quality factors of these eigenmodes are shown in Fig. 3B, cavity IV, of the main text.

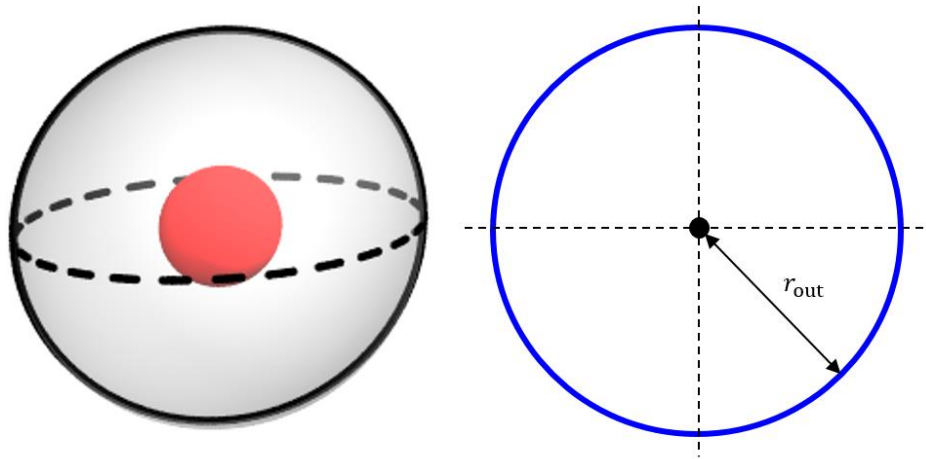

**Fig. S14.** Sketch (left) and dimensions (right) of the cavity studied in Figs. 4I to 4V of the main text. The cavity consists of a Si sphere ($\varepsilon_i = 11.7$) (shown in red) of radius $r_i = 1.507$ μm immersed in a SiC spherical host (shown as grey background) of different radii: 4I - $r_{\text{out}} = 3$ μm, 4II - $r_{\text{out}} = 4$ μm, III - $r_{\text{out}} = 5$ μm, IV - $r_{\text{out}} = 7.5$ μm and V - $r_{\text{out}} = 10$ μm. The location of the sphere in the main cavity is indicated by a black dot.

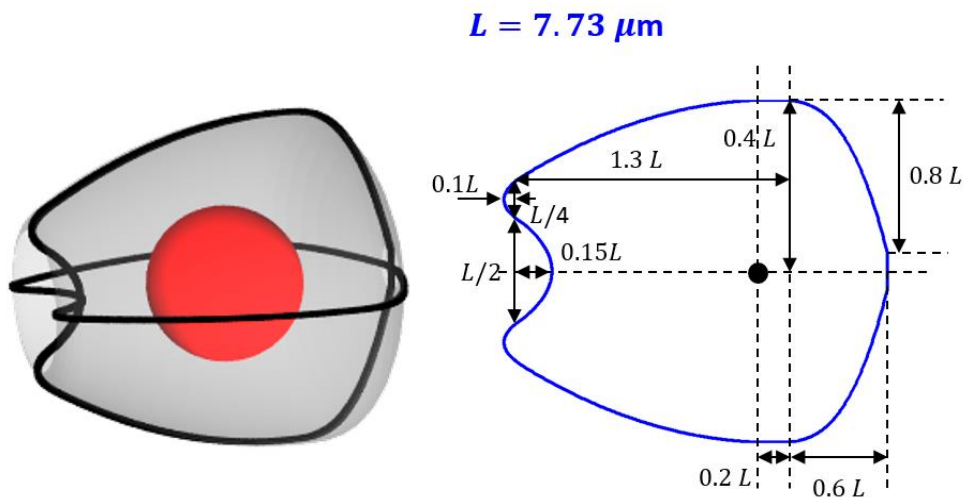

**Fig. S15.** Sketch (left) and dimensions (right) of the cavity studied in Fig. 4-VI of the main text. The cavity consists of a Si sphere ($\varepsilon_i = 11.7$) (shown as a red sphere) of radius $r_i = 1.507$ μm immersed in a SiC host (shown as grey background). The location of the sphere in the main cavity is indicated by a black dot.

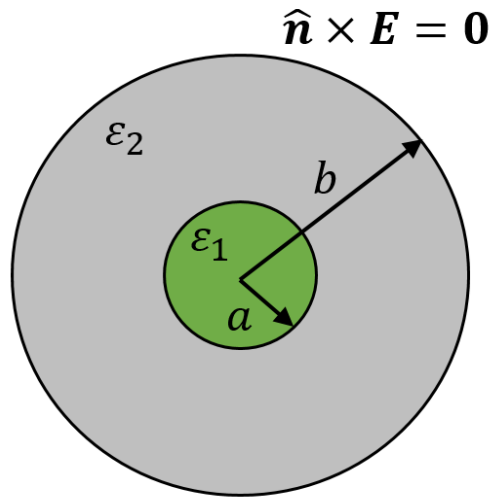

**Fig. S16.** Sketch of a cavity consisting of two concentric spheres, with internal and external radii equal to $a$ and $b$, respectively, and with internal and external permittivities equal to $\varepsilon_1$ and $\varepsilon_2$, respectively.

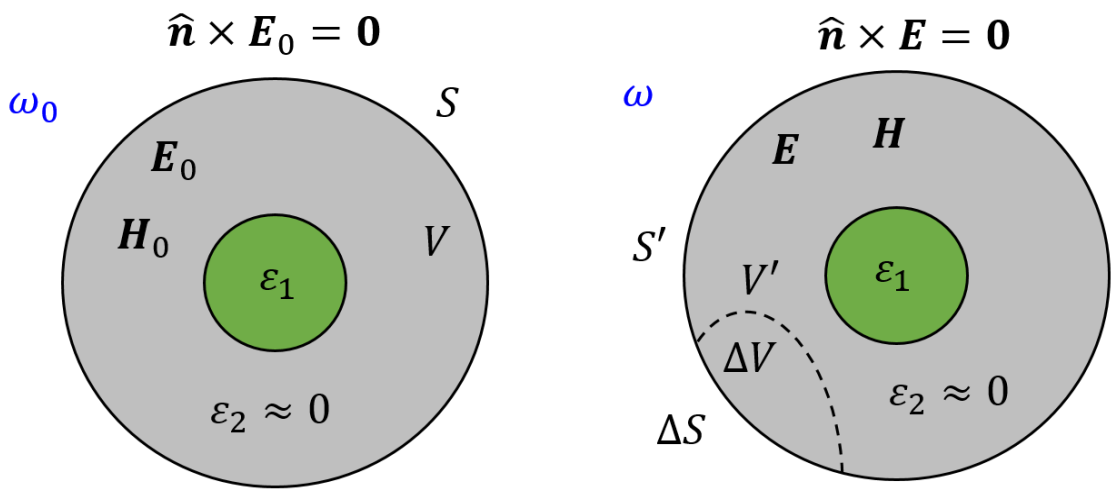

**Fig. S17.** (Left) Cavity characterized by volume $V$ and surface $S$, supporting and eigenmode with fields $\boldsymbol{E}_0, \boldsymbol{H}_0$ at the eigenfrequency $\omega_0$. (Right) Perturbed cavity with volume $V' = V - \Delta V$ and volume $S' = S - \Delta S$ supporting and eigenmode with fields $\boldsymbol{E}, \boldsymbol{H}$ at the eigenfrequency $\omega$.